\documentclass[prl,twocolumn,preprintnumbers,amsmath,amssymb,
floatfix,showpacs]{revtex4}
\usepackage{hyperref}
\usepackage{amsmath}
\usepackage[dvips]{epsfig}
\newcommand{\beq}{\begin{equation}}
\newcommand{\eeq}{\end{equation}}
\newcommand{\beqa}{\begin{eqnarray}}
\newcommand{\eeqa}{\end{eqnarray}}
\newcommand{\ba}{\begin{array}}
\newcommand{\ea}{\end{array}}

\begin{document}
\title{Kohn-Sham approach to Fermi gas superfluidity: 
the bilayer of fermionic polar molecules}

\author{Francesco Ancilotto$^{1,2}$} 
\affiliation{$^1$Dipartimento di Fisica e Astronomia "Galileo Galilei" and CNISM, 
                 Universit\`a di Padova, via Marzolo 8, 35122 Padova, Italy \\
             $^2$CNR-IOM Democritos, via Bonomea, 265 - 34136 Trieste, Italy} 

\begin{abstract} 
By using a well established 'ab initio' theoretical approach 
developed in the past to quantitatively study the
superconductivity of condensed matter systems, 
based on the Kohn-Sham Density Functional
theory, I study the superfluid properties and the BCS-BEC
crossover of two parallel bi-dimensional layers 
of fermionic dipolar molecules, where the
pairing mechanism leading to superfluidity 
is provided by the inter-layer coupling between dipoles.
The finite temperature superfluid properties of both 
the homogeneous system and one
were the fermions in each layer are confined 
by a square optical lattice are studied at half filling
conditions, and for different values of the strength 
of the confining optical potential.
The T=0 results for the homogeneous system are found to be
in excellent agreement with Diffusion Monte Carlo results.
The superfluid transition temperature in the BCS region 
is found to increase,
for a given inter-layer coupling, with the strength of the 
confining optical potential.
A transition occurs at sufficiently small interlayer
distances, where the fermions becomes localized within the
optical lattice sites in a square geometry 
with an increased effective lattice constant,
forming a system of localized composite bosons. 
This transition should be signalled 
by a sudden drop in the superfluid fraction of the system.
\end{abstract} 

\maketitle

\section{\bf I. INTRODUCTION}
\medskip

The study of superfluidity in Fermi systems is a very 
attractive area of research in the field of ultracold atoms because  
of the direct implications 
for superconductivity in solid-state materials  
as well as for nuclear and quark matter\cite{baba15}. One of the most relevant
experimental result in this field has been 
the realization, obtained by tuning the inter-particle interaction 
via the use of Fano-Feshbach resonances, of the crossover from the  
Bardeen-Cooper-Schrieffer (BCS) superfluid phase of loosely bound  
fermion pairs to the Bose-Einstein condensate (BEC) of tightly bound composite  
bosons \cite{zwe12,gio08}. 

Of particular relevance are the studies aimed at understanding 
the pairing of  
fermions in strongly interacting two-dimensional (2D) Fermi gases, a  
subject again of great importance for condensed matter physics 
in view of the  
not yet fully understood character of the corresponding mechanism in  
high-temperature (layered) superconductors.
While charge transport within layers plays an essential role 
in superconductivity of high T$_c$ material, the long-range 
nature of the interactions among the particles 
(electrons/holes) belonging to adjacent layers is 
believed to largely affect the value of the critical temperature.
A bilayer of fermionic particles interacting via long-range potential
thus represents an excellent platform to simulate the interplay between
these effects in the properties of high T$_c$ superconductors.

Recently, the creation of ultracold dipolar gases of fermionic
molecules with large intrinsic dipole moments
has been achieved\cite{kni08,zwi15}, opening the
way to explore the fascinating many-body physics of correlated 
Fermi systems associated with the long-range, anisotropic nature
of dipolar interaction between molecules\cite{bar08,bar12,li14},
which include topological superfluidity\cite{bru08,coo09}, interlayer pairing
between two dimensional systems and the formation of dipolar 
quantum crystals\cite{sie11} and possibly stripe phases\cite{mar12}.

Two-dimensional (2D) dipolar systems are of 
particular interest, since the lifetime of
heteronuclear molecules with permanent
electric dipole moment is increased by the effect of 
2D confinement\cite{dem11}. Indeed such polar molecules
can have very large dipole moments, of the order of 1 Debye, allowing to
access the regime of strong correlations in a controllable way.

When a bilayer of 2D fermionic dipoles is formed, where the dipoles in
each layer are 
aligned perpendicularly to the planes by an external field,
in spite of the repulsive interaction between fermions belonging to the
same layer, a superfluid behavior is nonetheless expected, 
the pairing among fermions being provided by
the attractive head-to-tail dipolar interaction
between fermions belonging to different layers, which
results in a two-body
bound state for any value of the bilayer separation
\cite{sim76,arm10,kla10}.
The resulting coupling is predicted to cause 
superfluid behavior at sufficiently 
low temperatures\cite{pik10,bar11,zin12,pot10}.
Moreover, a cross-over is expected by varying the
inter-layer distance, as the system 
evolves from the weak-coupling BCS regime of 
largely overlapping Cooper pairs to the strong-coupling BEC 
regime of composite bosons\cite{miy83,ran90}.
Additional interest in bilayers of fermions is due to the 
strong analogies with the physics of electron-hole bilayers in 
semiconductor heterostructures\cite{dep02}.
The BCS-BEC crossover in a (homogeneous) bilayer of fermionic dipoles
has been recently studied at zero-temperature by means of Diffusion
Monte Carlo (DMC) simulations\cite{gio14}. 

Density Functional Theory (DFT) for electrons, which is 
perhaps the most widely used and successful technique in 
electronic structure calculations of condensed matter 
systems, has been proposed only recently 
\cite{tro12,bulgac,gor15} as a useful computational tool in the field of cold gases.
The main advantage of the method is that it allows to go beyond the 
mean-field level by taking into account correlation 
effects, and thus represents a valid alternative to more 
microscopic (but also computationally more demanding) 
approaches such as Quantum Monte Carlo (QMC), especially 
for extended and/or inhomogeneous systems.

A modified DFT approach, which has been used 
to study the properties of a unitary Fermi gas\cite{bulgac},
is based on a functional form 
which exploits the scale invariance of the unitary regime.
DFT approaches have been used recently to describe a Fermi dipolar
system in various "single-orbital" approximations 
(Thomas-Fermi \cite{zyl15}, Thomas-Fermi-Dirac \cite{fan11},
Thomas-Fermi-von Weizsacker \cite{zyl13,zyl14}).
In Ref.\cite{abe14} a parameter-dependent 
DFT-LDA approach was used to study small
number of harmonically trapped fermions.

The well-known 
Kohn-Sham (KS) mapping\cite{kohn} of the many-body problem into a 
non-interacting one makes the DFT approach applicable in practice, 
either within the Local Density Approximation (LDA) 
or by including suitable gradient corrections.
The KS-DFT approach does not require adjustable parameters,
and thus belongs to the family of the so-called 'ab initio' methods well
known in the electronic structure community.
Recently, the KS-DFT method has been applied to cold atomic Fermi gases in 
optical lattices\cite{tro12,pila}, the unitary trapped Bose gas
\cite{anci2} and to the study of a rotating dipolar Fermi gas\cite{anc15}.

The extension of multi-orbital KS-DFT approaches to superfluids is a  
challenging goal, opening important new perspectives in the  
theory of cold quantum gases.
A formally exact generalization of "normal-state" DFT for 
condensed matter systems
which explicitly includes in its formulation 
the superconducting order parameter has been proposed
to describe solid-state BCS superconductors \cite{lud88}, 
and was found to be able to accurately predict
experimental properties of superconducting materials, 
especially for systems where a theory beyond 
simple BCS superconductivity is needed.

I will use here a method based on the approach described 
in Ref.\cite{lud88} to study the superfluid behavior of 
a homogeneous bilayer of (polarized) dipolar fermionic molecules,
for which accurate T=0 results exist to compare with,
obtained by using QMC simulations\cite{gio14}.
To underline the capabilities of the method to treat in particular
inhomogeneous systems (which are often difficult
to study using QMC methods), I will also calculate 
the finite-temperature superfluid properties of this
system in the presence of an additional external potential
which simulates an optical 2D square lattice acting on the dipoles in each 
layer. I will compute, throughout the BCS-BEC crossover, 
the normal-state density, 
the superconducting gap, the condensate fraction
and the superfluid transition temperature, and
their changes as a function of the interlayer distances
and the depth of the confining optical potential wells.
The exchange-correlation energy of the homogeneous
system, which is an essential ingredient of the KS-DFT method, 
will be provided by virtually exact Diffusion 
Monte Carlo calculations\cite{gio12,gio14}.

\section{\bf II. METHOD}
\medskip

The two-dimensional, spin-polarized dipolar Fermi
gas is characterized by the (intra-layer) interaction
$
V=\sum _{i<j}^N {d^2 \over |{\bf r}_i-{\bf r}_j|^3}.
$
Here $d$ is the electric dipole moment of an atom/molecule and 
${\bf r}_i,{\bf r}_j$ are coordinates in the 2D $x-y$ plane.
Being the dipole moments aligned parallel to the
z-axis, the pair potential is purely repulsive.
The range of the dipole-dipole interaction 
is characterized by the length $r_0=Md^2/\hbar ^2$,
$M$ being the particle mass.
The adimensional interaction strength characterizing the system
is $k_F r_0$ (where $k_F=\sqrt{4\pi n}$ 
is the Fermi wavevector of the 2D uniform system at a density $n $).

The inter-layer interaction is given by
\beq
\label{vint}
V_{IL}(r,\lambda)= d^2 {r^2-2\lambda^2  \over (r^2+\lambda ^2)^{5/2} }
\eeq
where $\lambda $ is the separation between the two layers
and $r$ is the in-plane distance between two dipoles belonging to different 
layers. At variance with the always repulsive intra-layer interaction, the 
potential $V_{IL}(r)$ is attractive for $r<\sqrt{2}\lambda $.

\subsection{\bf IIA. NORMAL STATE CALCULATIONS}
\medskip

Within the 
Kohn-Sham formulation\cite{kohn} of Density Functional 
Theory \cite{gross} for an inhomogeneous system of $N$ interacting  
particles with mass $M$, 
the total energy of the system is given by the
following total energy functional of the density $n$,
which includes the exact kinetic energy of a fictitious non-interacting
system and the interaction energy functional $E_{int}$:

\beq
E_{KS}[n({\bf r}) ]=-{\hbar ^2 \over 2M}\sum _i \int{\phi _i^\ast({\bf r}) 
\nabla ^2 \phi _i({\bf r})d{\bf r}} +E_{int}[n({\bf r}) ]
\label{ks}
\eeq

In the usual KS-DFT scheme for electronic systems, $E_{int}$ is usually split
into three contributions, i.e. the Hartree ("mean-field"), 
exchange and correlation terms.
The $\{\phi _i({\bf r}),\,i=1,N\}$ are single-particle orbitals,
forming an orthonormal set, $\langle \phi _i | \phi _j \rangle=\delta _{ij}  $.
I assume here a fully balanced system, 
with $N$ fermions per each layer. The total density of the system is
$n ({\bf r})=\sum _{i=1}^N |\phi _i({\bf r})|^2$.

In the present case a more convenient partition is $E_{int}[n ]
=E_{L}[n ]+E_{IL}[n ]$.
$E_{L}$ is the energy contribution due to 
intra-layer interactions, given by the sum of the
direct+exchange interaction term 
(the "Hartree-Fock" energy, $E_{HF}$)
and the correlation energy $E_C$, which I write here in the 
Local Density Approximation\cite{anc15}:

\begin{eqnarray}
\label{lda}
E_L[n]=E_{HF}[n]+E_C[n]\nonumber \\ =\int  
[ {256\over 45 } d^2 \sqrt{\pi }n ({\bf r})^{5/2}+
n ({\bf r}) \epsilon _C(n ({\bf r}))] d{\bf r}
\end{eqnarray}
where $\epsilon _C (n )$ is the correlation energy per particle 
of the {\it homogeneous} system of density $n $, as obtained
from the (virtually exact) Diffusion Monte Carlo
calculations of Ref.\cite{gio12}. 
The actual analytical form of the 
function $\epsilon _C (n )$ used to fit the DMC results 
is taken from Ref.\onlinecite{abe14}.

The inter-layer interaction energy $E_{IL}$ is given by the sum of the 
Hartree term plus the correlation energy (I neglect here any exchange
interaction contribution since orbitals of fermions on different layers
have zero overlap):

\beq
\label{lda1}
E_{IL}[n ]={1\over 2}\int d{\bf r}\int d{\bf r^\prime } 
n ({\bf r}) n({\bf r^\prime })  V_{IL}(|{\bf r}-{\bf r^\prime }|)
+\tilde {E}_C[n ]
\eeq

Note that, from Eq.(\ref{vint}), $\int V_{IL}(r) d{\bf r}=0$ \cite{bar11}, i.e. 
the mean-field interaction energy between two layers in the homogeneous case
($n ({\bf r})=n$) is zero. 
Corrections to the mean-field approximation for the 
inter-layer interaction energy are incorporated in the
correlation energy functional $\tilde {E}_C$.
Informations about this term come from the results 
of DMC calculations for the homogeneous bilayer system\cite{gio14},
where the corrections to mean-field results have been computed
as a function of $k_F\lambda$
(see Fig.(3) in Ref.\onlinecite{gio14}).
These corrections (once the energy of a single-layer dipolar Fermi liquid
, $E/N=0.6931\,\epsilon_F$ \cite{gio12},  
has been subtracted from the DMC results) can be written in the form:

\beq
\label{eneil}
\tilde {E}_{C}/N=-{\epsilon _F \over 2} f(k_F\lambda)
\eeq
where the function $f$ interpolates the DMC results.
I choose here $f(x)=0.292[1-\tanh[5.5(x-0.45)]]$, which gives a reasonable
overall fit to the DMC data.
The above result, which holds for the homogeneous system, can 
be used for inhomogeneous systems as well by 
using again the LDA:

\beq
\label{eneil1}
\tilde {E}_{C}[n]= -{1\over 2}\int d{\bf r}\,n({\bf r})\epsilon _F({\bf r}) f(k_F({\bf r})\lambda)
\eeq
where $k_F({\bf r})\equiv \sqrt{4\pi n({\bf r})}$ and 
$\epsilon ({\bf r})\equiv \hbar ^2k_F({\bf r})^2/2M$ are
the local values of the Fermi wave vector and energy, respectively.

The effective potential $\mu _{IL}({\bf r})\equiv {\delta E_{IL}\over \delta n}$ 
associated with the inter-layer interactions is thus

\begin{eqnarray}
\label{mu1}
\mu _{IL}({\bf r})
=\int d{\bf r^\prime}n({\bf r^\prime})V_{IL}(|{\bf r}-{\bf r^\prime}|)\nonumber \\
-{\epsilon _F({\bf r})\over 2}[2f(k_F({\bf r})\lambda+n({\bf r}) {\partial f\over \partial n }]
\end{eqnarray}

Constrained minimization of the energy functional $E_{KS}[n ]$
leads to the coupled KS eigenvalues equations
\begin{eqnarray}
 \label{kseq}
  [-{\hbar ^2 \over 2M }\nabla ^2 + V_{KS}({\bf r})]\phi _i({\bf r})=\epsilon _i \phi _i({\bf r})
\end{eqnarray}
where 
\begin{eqnarray}
 \label{ham}
  V_{KS}({\bf r})=\epsilon _C(n ({\bf r}))+n ({\bf r})
{\partial \epsilon _C \over \partial n}\nonumber \\
+{128\over 9}d^2\sqrt{\pi}n ^{3/2}({\bf r})+\mu _{IL}({\bf r})
\end{eqnarray}

Solutions of the above system of equations provide the density $n ({\bf r})$
(and thus the total energy, through Eq.(\ref{ks}))
of the fermion system in its (normal) ground-state.

In practice, the 
solutions $\{ \phi _i ({\bf r})\}$ of Eq.(\ref{kseq})
are found 
by propagating in imaginary time the time-dependent version\cite{kohn} of 
the KS equations (\ref{kseq}) (for more details about the actual method
used to efficiently propagate the orbitals $\phi _i$ in imaginary time, 
see Ref.\onlinecite{anc}).
Both the density and the single-particle orbitals $\phi _i$ have been
discretized in cartesian coordinates using
a spatial grid fine enough to guarantee
well converged values of the total energy. 
The orthogonality between different orbitals has been enforced
by a Gram-Schmidt (G-S) process. The spatial
derivatives entering Eq.(\ref{kseq}) have been calculated with 
accurate 13-point formulas, while
efficient Fast-Fourier techniques\cite{frigo} have been used to 
calculate the non-local term entering the KS potential $V_{KS}$
and the potential term entering the gap equation
(see the following Section).

\subsection{\bf IIB. SUPERFLUID STATE CALCULATIONS}
\medskip

The basic formulation of the KS-DFT for superconductors\cite{lud88},
which I will follow, {\it mutatis mutandis}, in the present work,
is described in the following.

The theory is based on the fermion (electron) density 
$n ({\bf r})=\langle\Psi ^+({\bf r}) \Psi ({\bf r})\rangle$ ("normal" density)
as well as the superconducting order parameter ("anomalous" density)
$\chi({\bf r},{\bf r}^\prime)=\langle \Psi ({\bf r}) \Psi ({\bf r^\prime})\rangle$
where $\Psi ^+({\bf r})$ is the fermion creation operator.
This quantity is finite for superconductors below the
transition temperature and zero above it.
Associated to these two densities there are two key quantities,
i.e. the KS potential $V_{KS}({\bf r})$ described in the 
previous Section,
and the so-called anomalous potential $\Delta _s({\bf r},{\bf r}^\prime)$:

\beq
\label{gapr}
\Delta _s({\bf r},{\bf r}^\prime)=\chi({\bf r},{\bf r}^\prime)V(|{\bf r}-{\bf
r}^\prime|)+\Delta _{xc}({\bf r},{\bf r}^\prime)
\eeq

Here $V(|{\bf r}-{\bf r}^\prime|)$ represents the effective interaction
between the fermionic particles responsible for pairing. In the present case 
$V\equiv V_{IL}$, where $V_{IL}$ is
the inter-layer dipole-dipole interaction potential, Eq.(\ref{vint}).
The first term in Eq.(\ref{gapr}) 
corresponds to the Hartree (mean-field) approximation,
while the extra term
include exchange and correlation effects. Although recipes have been
proposed to approximately construct $\Delta _{xc}$ for electronic 
superconductors\cite{lud88}, which may be adapted to the case of
fermionic cold gases, I neglect it here because
in the present case the attractive interaction acts between fermions
belonging to different, spatially separated 2D layers, and thus
exchange is null. I am nevertheless including 
neglected correlation effects beyond mean-field  
in the chemical potential (\ref{mu1}).

The Kohn-Sham Bogoliubov-de Gennes equations read\cite{bogo,lud88}

\beq
\label{eq1}
\big [ -{\nabla ^2  \over 2}  + V_{KS}({\bf r})-\mu     \big ]u_i({\bf r})+
\int d{\bf r^\prime } \Delta _s ({\bf r},{\bf r}^\prime) v_i({\bf r}^\prime)=
\tilde {E_i}u_i({\bf r})
\eeq

\beq
\label{eq2}
-\big [ -{\nabla ^2  \over 2}  + V_{KS}({\bf r})-\mu     \big ]v_i({\bf r})+
\int d{\bf r^\prime } \Delta _s ^\ast ({\bf r},{\bf r}^\prime) u_i({\bf r}^\prime)=
\tilde {E_i}v_i({\bf r})
\eeq
where $u_i({\bf r}), v_i({\bf r})$ are the particle and hole amplitudes.

We notice here that 
the non-local nature of the pairing field $\Delta _s ({\bf r},{\bf r}^\prime)$ 
in the above equations does not lead to 
the ultra-violet divergence in the 
anomalous density matrix elements which 
may occur\cite{bulgac1,bruun} 
when using, instead of Eq.(\ref{eq1},\ref{eq2}), the standard
Hartree-Fock Bogoliubov-de Gennes equations of the BCS mean-field theory of
superconductivity \cite{legget} with a local pairing field $\Delta _s({\bf r})$.
 
The amplitudes $u_i({\bf r}), v_i({\bf r})$ 
can be expanded in the complete set 
of wavefunctions $\{\phi _i({\bf r})\}$ of the normal-state 
Kohn-Sham equations:

\beq
 \label{kseq1}
  [-{\hbar ^2 \over 2M }\nabla ^2 + V_{KS}({\bf r})-\mu]\phi _i({\bf r})=
\epsilon _i \phi _i({\bf r})
\eeq

Within the so-called "decoupling approximation"\cite{lud88,kurth},
i.e. assuming
$u_i({\bf r})\sim u_i \phi _i({\bf r})$ and 
$v_i({\bf r})\sim v_i \phi _i({\bf r})$ 
(with $u_i$ and $v_i$ complex constants),
one can write  $\tilde {E}_i=\pm E_i$, where

\beq
\label{bige}
E_i=\sqrt{\xi _i^2 + |\Delta_i|^2}
\eeq
and
$\xi=\epsilon _i-\mu$

By defining the matrix elements:

\beq
\label{gapi}
\Delta _i = \int d{\bf r}\int d{\bf r}^\prime \phi _i^\ast({\bf r})
\Delta _s({\bf r},{\bf r}^\prime)\phi _i({\bf r}^\prime)
\eeq

one can write the following equations for the
normal and the anomalous densities:

\beq
\label{number}
n ({\bf r})={1\over 2}\sum _i [1-{\xi_i   \over E_i  }\tanh(\beta E_i/2)]|
\phi _i({\bf r})|^2
\eeq

\beq
\label{cai}
\chi ({\bf r},{\bf r}^\prime)={1\over 2}\sum _i {\Delta _i\over E_i}\tanh(\beta E_i/2)
\phi _i({\bf r})\phi _i^\ast ({\bf r}^\prime)
\eeq

Using Eq.(\ref{gapr}) and (\ref{gapi}) one can write Eq.(\ref{cai}) as an
implicit equation for $\Delta _i$:

\beq
\label{delta}
\Delta _i={1\over 2} \sum _j  {\Delta _j\over E_j}\tanh(\beta E_j/2)
\int d{\bf r}\phi _i^\ast ({\bf r})\phi _j^\ast ({\bf r})A_{ij}({\bf r})
\eeq
where

\beq
\label{matr}
A_{ij}({\bf r})\equiv \int d{\bf r}^\prime \phi _i ({\bf r}^\prime)\phi _j^\ast ({\bf
r}) V_{IL}(|{\bf r}-{\bf r}^\prime|)
\eeq

The convolution integrals appearing in $A_{ij}({\bf r})$, which 
are the most time consuming step in solving the gap equation
(\ref{delta}), are efficiently 
performed by using Fast Fourier transforms\cite{frigo}, knowing that 
the Fourier transform of $V_{IL}(r)$ is 
\beq
\label{vintq}
V_{IL}(q)=-d^2q e^{-\lambda q}
\eeq

The actual calculations of the superfluid quantities 
are performed as follows:
{\it (i)} first the ground-state density $n ({\bf r})$ and the 
effective potential $V_{KS}$ are found 
by self-consistently 
solving the Eq.(\ref{kseq1}) for the occupied states $\{\phi _i({\bf r}),i=1,N \}$;
{\it (ii)} a larger number $N^\prime $ (whose minimum value necessary for converged results 
depends on the effective coupling
between the layers, i.e. on the interlayer distance $\lambda $) 
of orbitals $\{\phi _i({\bf r}),i=1,N^\prime \}$ are calculated 
in the effective potential $V_{KS}$ obtained from the previous step;
{\it (iii)} the number density equation (\ref{number}) is solved for $\mu $
using the normalization condition $\int n({\bf r})d{\bf r}=N$;
{\it (iv)} the gap equation (\ref{delta}) is then 
solved iteratively to provide $\Delta _i$. 

The pairing gap $\Delta _0$ separating the normal to superfluid state
is finally found as 

$$
\Delta _0=min _{\{E_i \} }\Delta _i 
$$

In the BCS regime ($\mu>0$) the pairing gap $\Delta _0$
equals $\Delta _i(\epsilon _F)$.

Step {\it (ii)} in the sequence described above 
typically requires the (non self-consistent) calculations  
of a very large number (up to a few thousands) of empty states $\phi _i$. 
Most of the computer time during this step is spent in the G-S process. 
To expedite this time-consuming part of the calculations,
I employed a Block Gram-Schmidt orthogonalization 
procedure which can be recast using BLAS-3 level matrix-matrix
multiplication operations\cite{umari}, which can be efficiently 
performed using cpu-optimized mathematical libraries.
This allows to speed up the calculation by  
a factor between 4 to 5 with respect
to the time spent doing the conventional G-S iteration, which 
uses much less efficient BLAS-1 level operations.

From the calculated KS orbitals $\{\phi _i({\bf r}),i=1,N^\prime \}$ 
the condensate number of Fermi pairs can also be easily computed, 
being defined as follows:

\beq
\label{cond}
n_c=\int d{\bf r}\int d{\bf r}^\prime |\chi ({\bf r},{\bf r}^\prime)|^2 =
{1\over 4}\sum _i {|\Delta _i|^2\over E_i^2}\tanh ^2 (\beta E_i/2)
\eeq

\section{\bf III. RESULTS AND DISCUSSION}
\medskip

I will first discuss the case of the {\it homogeneous} bilayer of
dipolar fermions, both at T=0 and at finite temperatures: 
a comparison with the accurate DMC 
results at T=0 will allow to asses the accuracy of the 
method employed here, which will be used in the following Section to 
address the more complex case of the {\it inhomogeneous} bilayer system.

I assume in the following calculations $d=0.8$ Debye, which is 
appropriate to $K_{40}Na_{23}$ molecules in
the experimental realization of Ref.\cite{zwi15}.
The mass $M$ is that of a $K_{40}Na_{23}$ molecule.
The spatial range of the potential is thus given by
$r_0=Md^2/\hbar ^2\sim 0.6\,\mu m $.
The adimensional interaction strength characterizing the system
is $k_F r_0$.
I will consider here a fermion density such that 
$k_F r_0=0.5$ (a relatively weak value 
which can easily be achieved in experiments), which is the case 
studied in the T=0 DMC calculations reported in Ref.\onlinecite{gio12}
and Ref.\onlinecite{gio14}. These results
represent a solid benchmark with which the results
discussed in the present paper will be compared, at least for the 
homogeneous system at T=0.
For such value of $k_F$ the interparticle distance $\langle r\rangle$ 
is larger than the range of the interaction, 
being $\langle r\rangle/r_0\sim 3.6$ (dilute system).

\subsection{IIIA. HOMOGENEOUS SYSTEM}
\medskip

\begin{figure}
 \epsfig{file=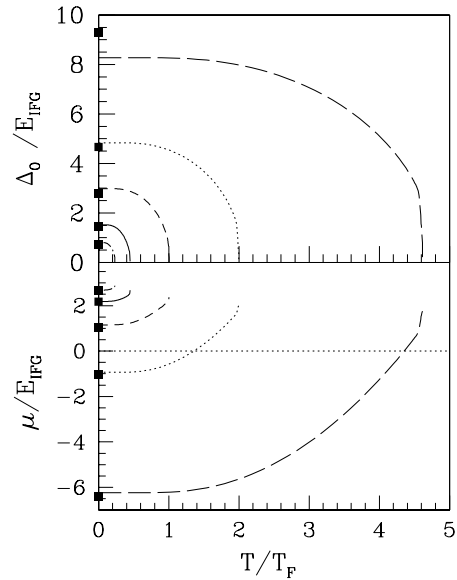,width=9 cm,clip=}
 \caption{Pairing gap $\Delta _0$ (upper panel) and chemical potential 
$\mu $ (lower panel) as a function of $T/T_F$, in units of
$E_{IFG}=\epsilon _F/2$, for different values of the interlayer distances:
$k_F\lambda = 0.25$ (dashed line),$\,0.3$ (dotted line),$\,0.35$ (short-dashed line),
$\,0.425$ (solid line) $\,0.5$ (dash-dot line). The squares at T=0 show the
DMC results from Ref.\onlinecite{gio14}.}
 \label{fig1}
\end{figure}

The calculated values of the pairing gap $\Delta _0$
and chemical potential $\mu $ for the 
homogeneous bilayer are shown, as a function of the 
temperature, in Fig.(\ref{fig1}). 
The temperature at which $\Delta _0=0$ is by 
definition the superfluid critical temperature $T_c$.
I compare these findings
at T=0 with the DMC results of Ref.\onlinecite{gio14}.
Note the excellent agreement throughout the whole
BCS-BEC crossover (which is conventionally set at 
$\mu =0$). The BEC regime is characterized by negative (large)
values of $\mu $, whereas in the BCS regime of weak coupling
$\mu >0$.

Although the agreement with the T=0 DMC results for the
chemical potential is not unexpected since two important ingredients 
in the effective potential (9) are fitted to 
DMC data (namely the inter- and intra-layer correlation 
contributions), the results for the pairing gap,
even at T=0, truly represent a prediction of the 
KS-DFT theory used here.
The calculated values for the gap show nonetheless a discrepancy 
for $k_F\lambda =0.25$, when compared with the
DMC result. I must recall, however, that according to Ref.\onlinecite{gio14}  
this value for the pairing gap have been computed 
using a different, approximate, expression
than that used for larger values of $k_F\lambda$.

From the calculated values of $\mu $ vs. $k_F\lambda$ at T=0 
I find that the
BCS-BEC crossing $\mu =0$ occurs at $k_F\lambda =0.322$, in almost perfect 
agreement with the DMC result, $k_F\lambda =0.325$.

\begin{figure}
\epsfig{file=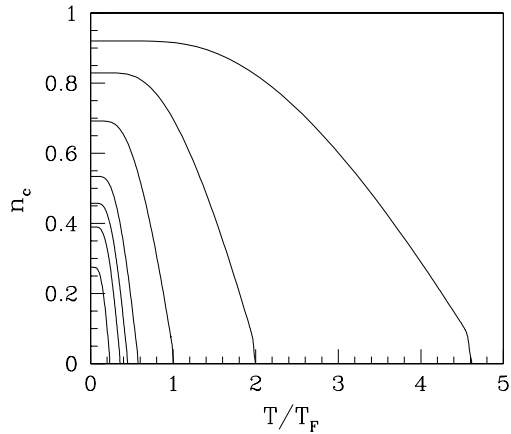,width=9 cm,clip=}
 \caption{Condensate fraction of the homogeneous bilayer, 
for different values of $k_F\lambda$.
From top to bottom: $k_F\lambda=0.25,0.3,0.35,0.4,0.425,0.45,0.5$}
 \label{fig2}
\end{figure}

In Fig.(\ref{fig2}) the calculated condensate fraction $n_c$ is shown
up to the superfluid critical temperature, for different values
of the interlayer distances. As expected, the condensate 
fraction decreases, as well as the critical temperature,
as the system evolves from BEC to BCS regime. 

\begin{figure}
 \epsfig{file=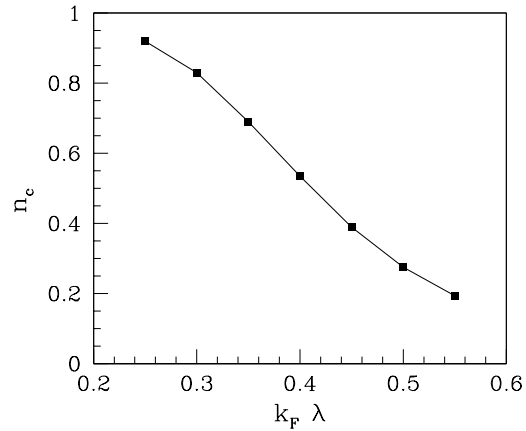,width=9 cm,clip=}
 \caption{Condensate fraction of the homogeneous bilayer at $T=0$  as a function
of the interlayer distance.}
 \label{fig3}
\end{figure}

The condensate fraction at T=0, calculated using Eq.(\ref{cond}), is shown in 
Fig.(\ref{fig3}) as a function of the interlayer distance.

\begin{figure}
 \epsfig{file=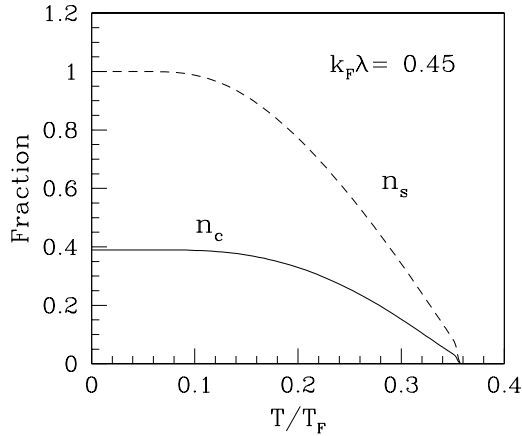,width=9 cm,clip=}
 \caption{Condensate (lower curve) and superfluid (upper curve) fraction as a function
of temperature. }
 \label{fig4}
\end{figure}

In the BCS region (i.e. for weak coupling resulting from
larger interlayer distance)  
the superfluid fraction $n_s$ can be calculated 
using Landau's formula\cite{fetter} for a two-dimensional system,
$n _s=n-n _n(T)$, where
the normal fluid component (assuming purely fermionic
excitations) is given by

\beq
\label{rhon}
n _n(T)={\hbar^2 \beta \over 2M} \int {d^2k \over (2\pi)^2}k^2 {e^{\beta E_k}\over
(e^{\beta E_k}+1)^2}
\eeq

$E_k$ represents the single-particle excitation spectra, which I
write here as $E_k=\sqrt{[\hbar ^2 k^2/2M^\ast-\mu]^2+\Delta _0 ^2}$,
using the effective mass $M^\ast$ as an adjustable parameter.
By imposing that the superfluid fraction
goes to zero at the calculated critical temperature $T_c$,
I find $M^\ast =0.7(0)M$, in reasonable agreement with the
effective mass $M^\ast =0.77\,M$ as computed in Ref.\onlinecite{gio14}
using DMC.
The calculated superfluid fraction, for $k_F\lambda =0.45$
(i.e. in the BCS regime), is compared in Fig.(\ref{fig4})
with the condensate fraction for the same interlayer distance.

Finally, the critical superfluid transition temperature 
is shown with a solid line in Fig.(\ref {fig7}).
It is known that in 2D the transition from normal to
superfluid state is of the Kosterlitz-Thouless (KT) type.
However, at least in the BCS limit, the KT transition temperature
is very close to the one calculated using BCS theory\cite{miy83}.
In the present case this is indeed true on the BCS side of the phase diagram.
I computed $T_{KT}$ through the Kosterlitz-Nelson condition:
\beq
\label{kt}
k_BT_{KT}={\hbar ^2 \pi \over 8m} n_s(T_{KT})
\eeq
and indeed found that it almost coincides with $T_c$ down to 
the $\mu =0$ line. 

\subsection{IIIB. INHOMOGENEOUS SYSTEM}
\medskip

I consider in the following a bilayer of 2D dipolar fermions
under the effect of an additional external potential in the
KS equations (\ref{kseq1})
corresponding to a square 2D optical lattice 

\beq
\label{opt}
V_{ext}({\bf r})=V_0[sin^2(x\pi/a)+sin^2(y\pi/a)]
\eeq
where two different well depths will be considered in the following, i.e.
$V_0=3.52\,\epsilon _F$ and $V_0=9.39\,\epsilon _F$. 
For the sake of simplicity, I will refer in the following 
to the first case as a "weak" potential, while
the second will be termed "strong".
The chosen lattice constant is $a=2.5\,k_F^{-1}$.

I will consider in the following 
the case of half-filling (i.e. 
one fermion every two lattice sites).

\begin{figure}
\epsfig{file=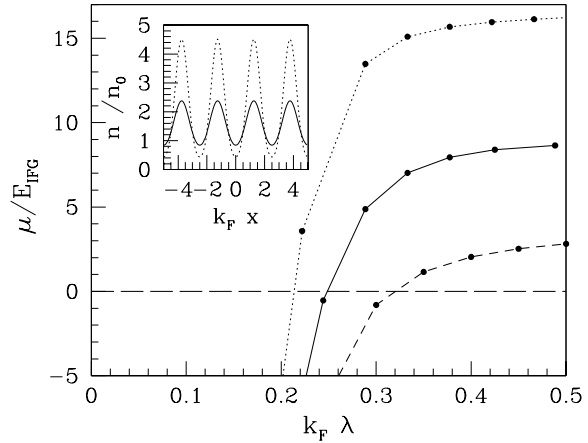,width=9 cm,clip=}
\caption{Chemical potential as a function of the interlayer distance, for the homogeneous
system (dashed line), the "weak" potential (solid line) and the "strong" 
potential (dotted line). The inset show the fermion density profiles, along the x-direction,
for the two optical potential strengths described in the text: $V_0=3.52\,\epsilon _F$
(solid line), $V_0=9.39\,\epsilon _F$ (dotted line).
Here $n_0$ is the density of the uniform system.}
\label{fig5}
\end{figure}

The chemical potential is calculated as described in 
Section.II as a function of the interlayer separation $\lambda $.
The results are shown in Fig.(\ref{fig5}), where they are compared with 
the results for the homogeneous bilayer treated in the 
previous Section. It appears that for stronger confining
potentials the BCS-BEC crossover moves towards lower values 
of the interlayer distance, i.e. it occurs for higher values
of the interlayer coupling, and the transition between BCS and BEC
regime becomes sharper.
The inset shows the fermion density profiles for the 
two values $V_0$ of the amplitude of the optical potential 
considered here, along a line 
passing through adjacent potential minima.

\begin{figure}
 \epsfig{file=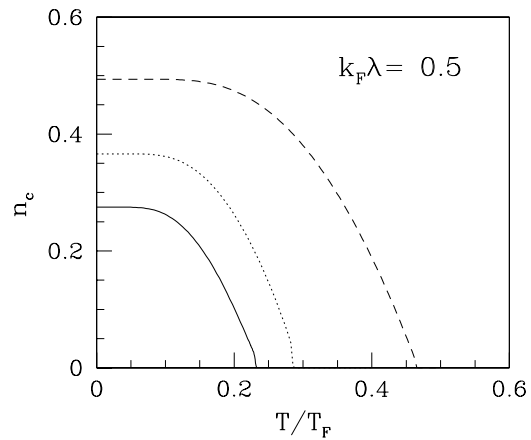,width=9 cm,clip=}
 \caption{Condensate fraction for the homogeneous bilayer system (solid line)
and for the two values of the strength $V_0$ of the optical potentials:
"weak" potential (dotted line) and "strong" potential (dashed line).
}
 \label{fig6}
\end{figure}

The condensate fraction is also shown, as a function of the
system temperature, in Fig.(\ref{fig6}), where it is compared with the
homogeneous bilayer results.

\begin{figure}
 \epsfig{file=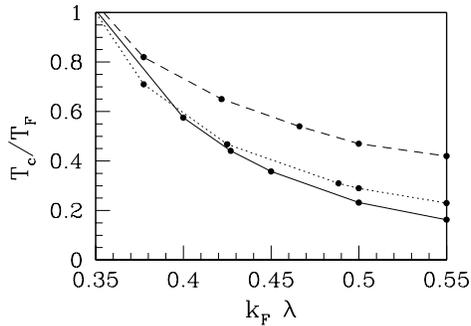,width=9 cm,clip=}
 \caption{Calculated critical temperatures for the homogeneous system 
(solid line) and for the two values of the strength $V_0$ of the optical potentials:
"weak" potential (dotted line) and "strong" potential (dashed line).  }
 \label{fig7}
\end{figure}

Finally, the dependence of the calculated 
critical superfluid temperatures on the interlayer distance is shown for the 
two optical potential strengths studied here (and for the homogeneous system as well)
in Fig.(\ref{fig7}).
Notice that in the BCS regime the critical temperature in the
presence of a strong optical potential is quite enhanced
with respect to the homogeneous system, the enhancement  
increasing as one goes deeper into the BCS regime. 
This finding is consistent with similar
results found for a different systems of fermionic atoms
subject to optical potentials, 
which undergo a phase transition to a superfluid 
state at a strongly increased transition temperature
with respect to the uniform case\cite{hof02}.

I end this Section by mentioning a transition which takes
place in the presence of the "weak" optical potential, 
when the distance between the two dipolar layers is reduced 
until it reaches a critical (small)
value $\lambda _c$, corresponding to a large and 
negative value of $\mu $, i.e. in the deep BEC regime. 
Then there occurs a sudden rearrangements of the 
orbital occupations, so that the fermions that for $\lambda > \lambda _c$
populate each lattice site with 1/2 filling,
become suddenly localized every other lattice site 
(which become populated by one fermion each)
as the critical interlayer distance $\lambda _c$ is reached.

\begin{figure}
 \epsfig{file=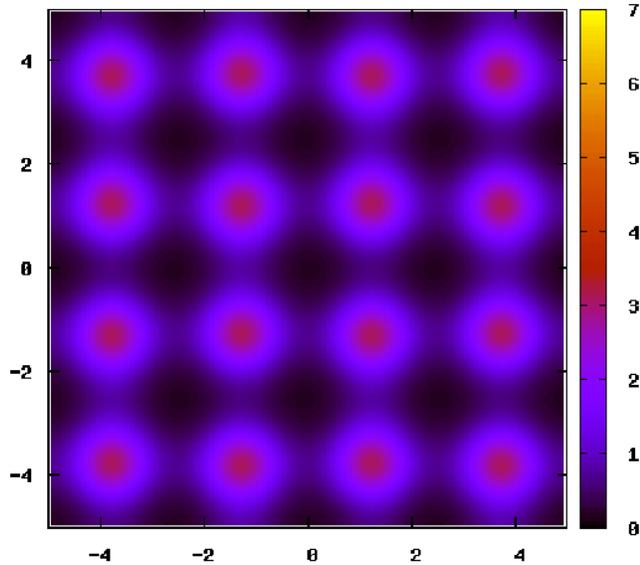,width=9 cm,clip=}
 \caption{(Color online) Contour plot of the total density 
for $k_F\lambda = 0.12$ in the "weak" 
optical potential. The coordinates are in units of $k_F^{-1}$,
while the density is in units of $n_0$.
}
 \label{fig8}
\end{figure}

\begin{figure}
 \epsfig{file=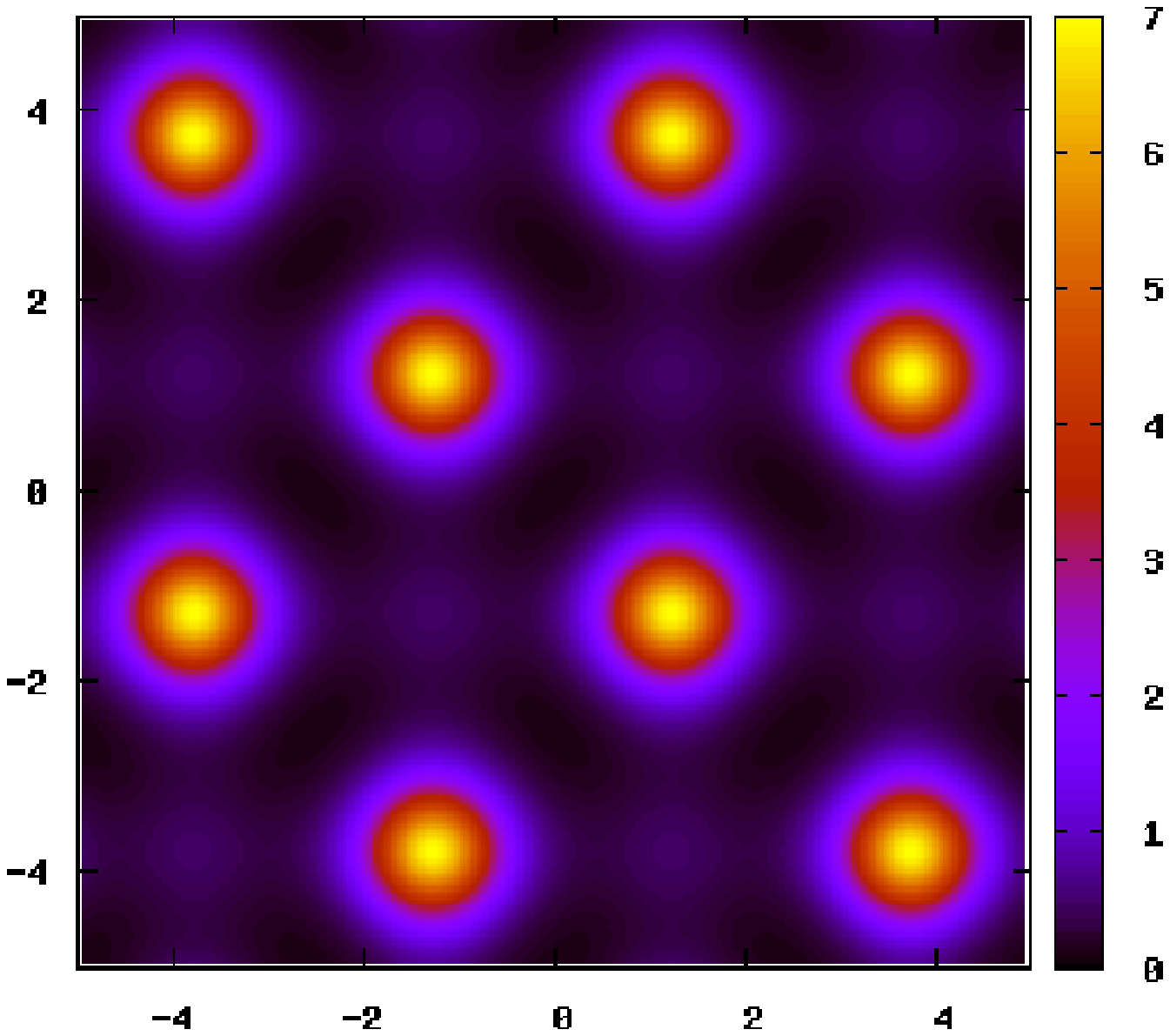,width=9 cm,clip=}
 \caption{(Color online) Contour plot of the total density for 
$k_F\lambda = 0.14$ in the "weak" optical potential. 
The coordinates are in units of $k_F^{-1}$,
while the density is in units of $n_0$.
}
 \label{fig9}
\end{figure}

\begin{figure}
 \epsfig{file=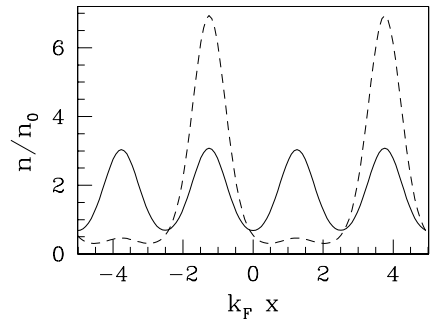,width=9 cm,clip=}
 \caption{Density profiles along the x-direction, corresponding to the 
two structures shown in Fig.(\ref{fig8}) and Fig.(\ref{fig9}). 
$n_0$ is the density of the homogeneous system.}
 \label{fig10}
\end{figure}

This transition (which occurs at $k_F\lambda_c =0.13$) 
is illustrated by the associated fermion density
in the optical lattice immediately before and 
after it, in Fig.(\ref{fig8}) and Fig.(\ref{fig9}), respectively.
The sudden localization of each fermion in every other lattice site
results in a marked change in the total fermion density,
which suddenly evolves from a "delocalized" configuration characterized by the 
lattice constant $a$ (see Fig.(\ref{fig8})) 
to a $45^\circ $-rotated square lattice structure
with a larger lattice constant, $a^\prime =\sqrt{2}a$ (see Fig.(\ref{fig9})).
In this "localized" phase, each dipolar fermion sits in
a lattice site just above an equally occupied one in the
second layer, thus forming a tightly
bound composite boson because of the dipole-dipole
head-to-tail attractive interaction. 
In both phases shown in Fig.(\ref{fig8}) and Fig.(\ref{fig9}), 
the calculated condensate fraction is close to 1.
This sudden transition is not accompanied by 
any visible anomaly or discontinuity neither in the chemical
potential nor in the superconducting gap or the condensate fraction
as a function of $k_F\lambda$. However, I speculate that
the resulting system of localized composite bosons 
is expected to be a superfluid, as a result of the 
finite overlap between neighboring sites, although the superlfuid
fraction is expected to be smaller than in the 1/2 filling phase
shown in Fig.(\ref{fig8}).
This is suggested in Fig.(\ref{fig10}) where the density profiles 
along the x-axis are shown for the two structures 
in Fig.(\ref{fig8}) and Fig.(\ref{fig9}).
Due to the reduced overlap between fermion pairs
in the "localized" phase, 
one should expect, once the condition for this
transition is met by varying the interlayer distance, 
to observe a sudden drop in the superfluid fraction.
In a way, the resulting state, being characterized
by strong density modulations and a 
finite superfluid fraction, shares some similarities with 
a truly "super-solid" phase.

To substantiate quantitatively the above speculations
the actual superfluid fraction in both phases
shown in Fig.(\ref{fig8}) and Fig.(\ref{fig9})
should be explicitly computed. 
The superfluid fraction could be extracted, 
for instance, from the calculated total momentum
of the system under a Galilean boost, as obtained  
by solving the associated time-dependent equations of motion 
in the co-moving frame of reference
(as done, for example, in Ref.\onlinecite{soft} for a system of 
soft-core bosons).
Since this is not the main subject of the present paper, I
will not do this here.

In a practical, quasi-2D realization of the 
bilayer geometry studied here the transition
discussed above, which
involves a sudden evolution of the system from an
"itinerant" character associated with the first structure, 
to a more "insulating" one for the second structure, 
should be observable in principle by studying the
transport properties of the bilayer system trapped inside
an optical lattice with underlying harmonic confinement
(see, for instance, Ref.\onlinecite{str07}), e.g. by monitoring
the center of mass motion of the atomic cloud after a sudden displacement
of the harmonic trap minimum.          
As discussed in Ref.\onlinecite{str07}, systems with high filling 
are characterized by a slower relaxation towards the equilibrium position.
Accordingly, the decrease in the superfluid fraction expected 
with the transition described here should be signalled by a discontinuity
in the observed relaxation time of the displaced clouds.

\section{\bf IV. CONCLUSIONS}
\medskip

In this paper I applied a  
multi-orbital Kohn-Sham 'ab initio' formulation of the Density Functional
Theory of BCS superconductivity in condensed matter system, 
which is known to accurately predict the
experimental properties of superconducting materials
especially for systems where a theory beyond
simple BCS superconductivity is needed,
to the study of Fermi gas superfluidity, 
namely in a bilayer of fermionic dipolar molecules in 2D,
aligned perpendicular to the planes,
where the superfluid pairing is provided by the partially 
attractive interaction between dipoles belonging to different layers.

The finite temperature superfluid properties of both 
the homogeneous system and one
where the fermions in each layer are confined 
by a square optical lattice have been studied.
The resulting T=0 properties of the 
homogeneous system are found to be in excellent agreement with 
the results of recent Diffusion Monte Carlo simulations.

I computed the superconducting gap, the condensate fraction
and the superfluid transition temperatures, and found how they 
change as the interlayer distances is varied,
together with their dependence upon the depth of 
the confining optical potential wells.
A marked increase in the superfluid critical 
temperature in the BCS regime is found with increasing 
amplitude of the confining potential.

When the distance between the two dipolar layers reaches a critical
(small) value, corresponding to coupling strengths
characteristic of the deep BEC regime,
a transition is observed where the fermions, previously
spread out over the lattice, increase their localization 
such as there is one composite boson in every other lattice site.
This transition should be signalled by a sudden drop in the system
superfluid fraction, due to the reduced overlap 
between neighboring particles. 

\medskip
\medskip
{\bf Acknowledgments}
\medskip

The author thanks Luca Salasnich, Giacomo Bighin, 
Flavio Toigo, Paolo Umari and Pier Alberto Marchetti
for useful discussions and comments.

\end{document}